\documentclass[%
 reprint,
 amsmath,amssymb,
 aps,
]{revtex4-2}

\usepackage{graphicx}
\usepackage{dcolumn}
\usepackage{bm}
\usepackage{xcolor}
\usepackage{hyperref}

\begin{document}

\preprint{APS/123-QED}

\title{Sympathetic Mechanism for Vibrational Condensation Enabled by Polariton Optomechanical Interaction}

\author{Vladislav~Yu.~Shishkov\textsuperscript{1,2}}

\author{Evgeny~S.~Andrianov\textsuperscript{1,2}}

\author{Sergei~Tretiak\textsuperscript{3,4}}

\author{K.~Birgitta~Whaley\textsuperscript{5,6}}

\author{Anton~V.~Zasedatelev\textsuperscript{7}}
\email{anton.zasedatelev@univie.ac.at}

\affiliation{
 \textsuperscript{1}Dukhov Research Institute of Automatics (VNIIA), 
 22~Sushchevskaya, Moscow~127055, Russia
}
\affiliation{
 \textsuperscript{2}Moscow Institute of Physics and Technology, 9~Institutskiy~pereulok, Dolgoprudny~141700, Moscow~region, Russia
}

\affiliation{
 \textsuperscript{3}Center for Integrated Nanotechnologies,~ Los~ Alamos~National~Laboratory,~ Los~Alamos,~ NM,~ 87545~ USA
}
\affiliation{
 \textsuperscript{4}Theoretical Division,~ Los~ Alamos~ National~ Laboratory,~ Los~ Alamos,~ NM,~ 87545~ USA
}

\affiliation{
 \textsuperscript{5}Department of Chemistry,~ University~ of~ California,~ Berkeley,~ California ~94720,~ USA
}
\affiliation{
\textsuperscript{6}Berkeley Center for Quantum Information and Computation,~ Berkeley,~ California~ 94720,~ USA
}

\affiliation{
\textsuperscript{7}Vienna Center for Quantum Science and Technology~(VCQ),
~Faculty~of~Physics,~University~of~Vienna, Boltzmanngasse~5, 1090~Vienna, Austria
}

\date{\today}

\begin{abstract}

We demonstrate a macro-coherent regime in exciton-polariton systems, where nonequilibrium polariton Bose--Einstein condensation coexists with macroscopically occupied vibrational states.
Strong exciton-vibration coupling induces an effective optomechanical interaction between cavity polaritons and vibrational degrees of freedom of molecules, leading to vibrational amplification in a resonant blue-detuned 
configuration.
This interaction provide a sympathetic mechanism to achieve vibrational condensation with potential applications in cavity-controlled chemistry, nonlinear and quantum optics.

\end{abstract}

\maketitle

The formation of new eigenstates - cavity polaritons - through strong light-matter interaction, grants remarkable control over the physical and chemical properties of molecular materials. 
This opens new ways for the modification of chemical reactions~\cite{hutchison2012modifying,ahn2023modification,yuen2019polariton,garcia2021manipulating}, long-range energy transfer~\cite{saez2018organic,xiang2020intermolecular,georgiou2021ultralong}, steering of singlet/triplet dynamics~\cite{martinez2018polariton,martinez2019triplet,eizner2019inverting,polak2020manipulating}, and enhancement of nonlinear optical response~\cite{zasedatelev2019room,zasedatelev2021single}. 
Macroscopic quantum phenomena, such as nonequilibrium polariton Bose--Einstein condensation (BEC)~\cite{plumhof2014room,cookson2017yellow} and superfluidity~\cite{lerario2017room} in various organic materials at room temperature, have established light-matter BECs as a versatile platform suitable for diverse applications~\cite{keeling2020bose,jiang2022exciton,kavokin2022polariton,luo2023nanophotonics}.
Over the past decade, progress in exciton-polariton BEC has given rise to novel architectures in molecular optoelectronics, which include energy-efficient tunable coherent light sources not reliant on population inversion~\cite{sannikov2019room,wei2021low,putintsev2020nano}, ultra-fast all-optical transistors and logic gates~\cite{zasedatelev2019room,baranikov2020all,de2023room}, extreme photon nonlinearities for emerging quantum technologies~\cite{zasedatelev2021single}, and room-temperature topological states immune to disorder~\cite{dusel2021room,scafirimuto2021tunable}. 

While exciton-polariton BECs are relatively well explored, vibrational polariton condensation remains experimentally elusive. If realized, it will bring molecular vibrations to a macroscopic quantum state, where all vibrations share the same quantum properties, become indistinguishable and behave as of macroscopic coherent matter wave demonstrating long range order. However, achieving this state is challenging. Factors such as fast vibrational relaxation, relatively high thermal fluctuations at room temperatures, lack of appropriate experimental methods, and insufficient quality of cavity structures stand as current obstacles on a way towards vibrational BEC. Amidst these challenges, the significance of vibrational condensates has grown, particularly due to the recent progress in cavity-controlled chemistry~\cite{garcia2021manipulating}. Initially proposed in early 2000s, by Heinzen et al ~\cite{heinzen2000superchemistry} and Moore and Vardi~\cite{moore2002bose}, the idea of Bose-enhanced chemistry (or \textit{superchemistry}) exploits quantum statistics to accelerate chemical reactions for ultracold atoms, steering them into the product state — a quantum degenerate state of the BEC. Similarly, the genuine quantum statistics of vibrational condensates provides remarkable energetic and entropic advantages in the pathways of chemical reactions~\cite{pannir2022driving} with potential applications in complex molecular systems at room temperature. In this study, we investigate vibrational states in polariton systems with strong exciton-vibration coupling. We identify a new regime, characterized by macroscopically occupied vibrational states that coexist with a well-controlled exciton-polariton BEC. 
Our research introduces novel approaches to achieve vibrational condensation and could serve as a guideline for future experiments, taking advantage of strong exciton-vibration coupling as an inherent mechanism and general feature of molecular systems.


\begin{figure}[!ht]
\includegraphics[width=0.85\linewidth]{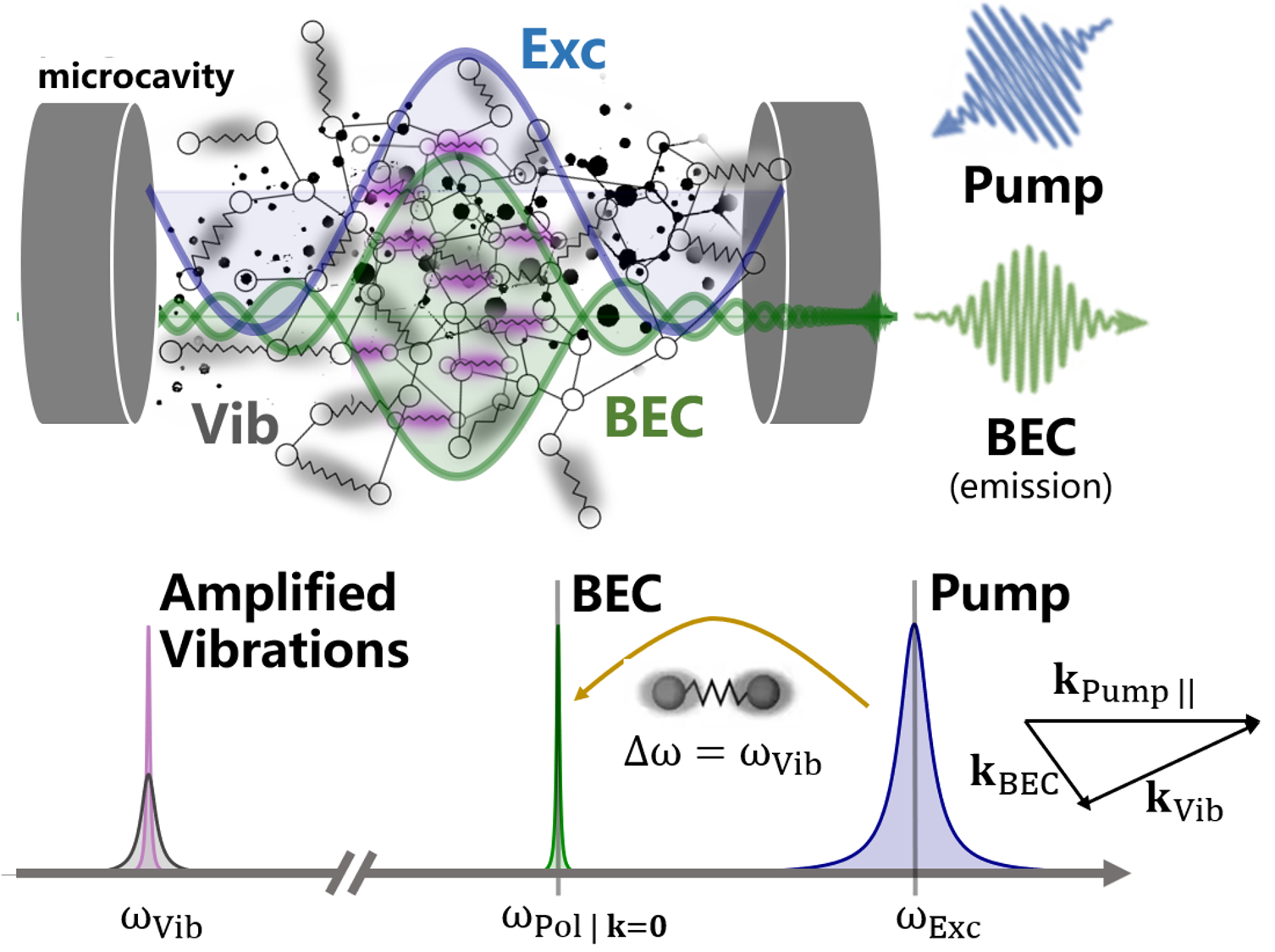}
\caption{Schematic of the setup to generate macroscopic vibrational states within exciton-polariton BEC in a strong vibronic regime. A coherent laser drive (Pump) in the blue-detuned optomechanical configuration resonantly excites bright excitonic states (Exc), causing vibrational amplification through optomechanical coupling between molecular vibrations (Vib) and the cavity polariton BEC satisfying the phase-matching condition: $\hbar \bf k_{\rm Pump||} = \hbar \bf k_{\rm BEC} + \hbar \bf k_{\rm Vib}$.}
    \label{fig:Fig1}
\end{figure}

Here we consider a nonequilibrium microscopic model that describes a large ensemble of organic molecules, each hosting a single exciton (${\hat H_{\rm Exc} = \sum_{j=1}^{N_{\rm mol}} 
{\hbar {\omega_{{\rm{exc}}}}\hat \sigma_{{\rm{Exc}}{j}}^\dag {{\hat \sigma}_{{\rm{Exc}}{j}}}}}$, where $N_{\rm mol}$ is the total number of molecules, $\omega_{\rm exc}$ is the eigenfrequency of the excitons, $\hat \sigma_{{\rm{Exc}}{j}}^\dag $ ($\hat \sigma_{{\rm{Exc}}{j}}$) is the creation (annihilation) operator for an exciton of a molecule located at the point ${{\bf r}_j}$) coupled strongly to an optical cavity (${\hat H_{\rm Cav} = \sum_{\bf k} 
\hbar {\omega_{{\rm Cav}|{\bf k}}}
\hat a_{{\rm Cav}|{\bf k}}^\dag {{\hat a}_{{\rm Cav}|{\bf k}}}}$, where $\hat a_{{\rm Cav}|{\bf k}}^\dag$ ($\hat a_{{\rm Cav}|{\bf k}}$) is the creation (annihilation) operator for a cavity photon with in-plane wavevector ${\bf k}$ and frequency ${\omega_{{\rm Cav}|{\bf k}}}$) and interacting with molecular vibrations (${\hat H_{\rm Vib} = \sum_{j=1}^{N_{\rm mol}} 
{\hbar {\omega_{\rm{Vib}}}\hat b_{{\rm Vib}j}^\dag {{\hat b}_{{\rm Vib}j}}}}$, where $\hat b_{{\rm Vib}j}^\dag$ ($\hat b_{{\rm Vib}j}$) is the creation (annihilation) operator of a molecular vibration with the eigenfrequency $\omega_{\rm Vib}$ for the corresponding molecule). The electronic excitation, initially confined to a single molecule, becomes delocalised through interaction with the cavity field: $\hat H_{\rm Exc-Cav} = \sum_{j=1}^{N_{\rm mol}} 
\sum_{{\bf k}} 
{\hbar \Omega_{j{\bf k}}
\left( 
\hat \sigma_{{\rm Exc}{j}}^\dag {{\hat a}_{{\rm Cav}|{\bf k}}}{e^{i{\bf k}{\bf r}_j}} 
+
h.c.
\right)}$, where $\Omega_{j{\bf k}}$ is Rabi frequency of the $j$-th molecule and cavity mode with wavevector $\bf k$~\cite{scully1997quantum}. This interaction implicitly involves exciton-vibration coupling via the matter component, represented by the term in the Hamiltonian: $\hat H_{\rm Exc-Vib} = \sum_{j=1}^{N_{\rm mol}} 
{\hbar \Lambda \omega_{\rm Vib}\hat \sigma_{{\rm Exc}j}^\dag {{\hat \sigma }_{{\rm Exc} {j}}}\left( {{{\hat b}_{{\rm Vib}j}} + \hat b_{{\rm Vib}j}^\dag } \right)}$, where $\Lambda$ is the interaction constant between excitons and molecular vibrations being a square root of the Huang--Rhys (HR) factor~\cite{atkins2011molecular}. It quantifies interaction between the electronic structure of a molecule and the displacement of its nucleus, which subsequently extend significant influence on the polaritonic degrees of freedom~\cite{herrera2018theory}. The full system Hamiltonian thus includes molecular excitons along with vibrational eigenenergies, cavity modes, and the aforementioned coupling terms between subsystems~\cite{shishkov2024optomech}
\begin{multline}\label{Full Hamiltonian simple form}
\hat H 
= 
\hat H_{\rm Exc}
+
\hat H_{\rm Cav}
+
\hat H_{\rm Vib}
+
\hat H_{\rm Exc-Cav}
+
\hat H_{\rm Exc-Vib}
.
\end{multline}

Considering the strong interaction between excitons, the cavity, and molecular vibrations, we necessitate a transformation to the momentum representation of the dressed states (see Ref.\cite{shishkov2024optomech} for details). The transformation unfolds in several stages. Firstly, we introduce vibrationally dressed excitons and dressed molecular vibrations~\cite{excOperator}. Subsequently, we account for the light-matter interaction of the dressed excitons with the cavity, yielding lower and upper polariton states~\cite{polOperator}. Excitons accessible by a dipole coupling from the ground state, termed ``bright excitons'', contrast with the remaining ``dark excitons''. Bright excitons and polaritons are phase-coherent, many-body delocalized states with a well-defined in-plane momentum $\hbar\bf k$, matching eigenstates of the cavity and in-plane component of the incident wavevector of a pump beam $\bf k_{\rm Pump ||}$. Dark excitons, lacking well-defined momentum, represent a manifold of localized states. The interplay between the dark and bright states significantly influences molecular dynamics in optical cavities, underpinning the ubiquitous $\sim1/N$ factor in rates of various population transfer processes~\cite{perez2023simulating}.
Analogously, we separate the dressed molecular vibrations into the bright and dark types. The bright vibrations are coherent and delocalized states. Similar to the bright excitons they have a well-defined momentum $\hbar\bf k_{\rm Vib}$~\cite{BvibOperator}, unlike the localized dark vibrational states.


Moving toward dynamic description of the system, we employ Lindblad approach to account for environmental interactions~\cite{carmichael2009open, scully1997quantum, breuer2002theory}. We establish a theoretical framework to calculate observable averages, including polariton, exciton, and vibration occupation numbers. Environmental interactions lead to decoherence necessitating an analysis of the relaxation and thermalization processes. For excitons, intermolecular interaction and non-radiative decay are key mechanisms~\cite{micha2007quantum,ostroverkhova2016organic}. However, for polaritonic states, cavity decay predominantly affects energy and phase loss in practical devices~\cite{keeling2020bose}. The coherence of vibrational states is typically lifetime-limited by a fast inter- and intramolecular vibrational energy redistribution (IVR)~\cite{martin2004femtochemistry}. Thermalization is another essential process for polariton BEC~\cite{bloch2022non}. In organic polariton systems, thermalization can occur due to intermolecular energy relaxation and via nonlinear interaction with low-frequency vibrations~\cite{tereshchenkov2023thermalization}. These processes are described by Lindblad superoperators $\hat L$ acting on the density matrix operator $\hat \rho$ (see details in Ref.\cite{shishkov2024optomech}).

We consider laser excitation of the bright excitonic states with in-plane momentum $\hbar\bf k_{\rm ex} = \hbar\bf k_{\rm Pump||}$, adhering to the resonant blue-detuned configuration $ \hbar\omega_{\rm Pump} = \hbar\omega_{\rm Exc} = \hbar\omega_{\rm Pol|\bf k = \bf 0} + \hbar\omega_{\rm Vib}$ 
as illustrated in Figure~\ref{fig:Fig1}. Here, $ \hbar\omega_{\rm Pump}$ and $\hbar\omega_{{\rm Pol}|\bf k = \bf 0}$ denote photon energy of the pump and the ground polariton state energy. Next, we derive a master equation and develop a mean-field theory for the system to simulate the dynamics of the average occupation numbers (see Supplementary Materials (SM)~\cite{sm} Section I, which includes~\cite{banyai2002real, kasprzak2006bose, sarchi2007long, doan2008coherence, guha2003temperature, coropceanu2002hole}). Going beyond the mean field, for instance by developing truncated Wigner theory~\cite{phuc2024semiclassical}, might provide additional insights into quantum statistics and possible entanglement generation between polariton and vibrational states. For the purpose of the study we restrict ourselves to the mean-field approach. The time evolution of average occupation numbers within the system is described by the equations below, and further elaborated in Ref.\cite{shishkov2024optomech}.

\begin{multline} \label{NumberOfBrightExcitons}
\frac{d n_{\rm Exc|{\bf k}_{\rm ex}}}{dt} 
=  
-  \gamma_{\rm Exc} ( n_{\rm Exc|{\bf k}_{\rm ex}} - \varkappa_{\rm Pump} )
+ 
\gamma_{\rm Exc}^{\rm B-D} ( n_{\rm Exc_{D}} - 
\\
n_{{\rm Exc}|{\bf k}_{\rm ex}} ) 
-
\sum_{\bf k} \frac{G_{{\bf k}}}{N_{\rm mol}} 
\left[
n_{{\rm Exc}|{\bf k}_{\rm ex}} \left(n_{{\rm Pol}|{\bf k}} + 1 \right) 
+
\right.
\\
\left.
n_{{\rm Vib}|{\bf k_{\rm ex}}-{\bf k}} \left( n_{{\rm Exc}|{\bf k}_{\rm ex}} - n_{{\rm Pol}|{\bf k}_{\rm ex}} \right) 
\right]
\end{multline}
\begin{multline} \label{NumberOfDarkExcitons}
\frac{d n_{\rm Exc_{D}}}{dt} 
=  
-  \gamma_{\rm Exc} n_{\rm Exc_{D}}
+ { \gamma_{\rm Exc}^{\rm B-D} \over N_{\rm mol}} ( n_{{\rm Exc}|{\bf k}_{\rm ex}} - n_{\rm Exc_{D}} ) 
-
\\
\sum_{\bf k} \frac{G_{\bf k}}{N_{\rm mol}} 
\left[
n_{\rm Exc_{D}} \left(n_{{\rm Pol}|{\bf k}} + 1 \right) 
+
n_{\rm Vib_{D}} \left( n_{\rm Exc_{D}} - n_{{\rm Pol}|{\bf k}} \right) 
\right]
\end{multline}
\begin{multline} \label{NumberOfLowerPolaritonsH}
\frac{d n_{{\rm Pol}|{\bf k}}}{dt} 
=  
-  \gamma_{{\rm Pol}|{\bf k}} n_{{\rm Pol}|{\bf k}}
+ 
G_{\bf k}
\left[
n_{\rm Exc_{D}} \left(n_{{\rm Pol}|{\bf k}} + 1 \right) 
+
 \right.
\\
\left.
n_{\rm Vib_{D}} \left( n_{\rm Exc_{D}} - n_{{\rm Pol}|{\bf k}} \right) 
\right]
+ 
{G_{\bf k} \over N_{\rm mol} } 
\left[
n_{\rm Exc|\bf {k_{ex}}} \left(n_{{\rm Pol}|{\bf k}} + 
\right.
\right.
\\
\left.
\left.
1 \right) 
+
n_{{\rm Vib}|{\bf k_{\rm ex}}-{\bf k}} \left( n_{{\rm Exc}|{\bf k}_{\rm ex}} - n_{{\rm Pol}|{\bf k}} \right) 
\right]
 + 
\sum_{\bf k'} \left\{ \gamma _{\rm therm}^{\bf kk'}
\right.
\\
\left.
\left( n_{{\rm Pol}|{\bf k}} + 
1 \right){n_{{\rm Pol}|{\bf k'}}} 
 - 
\gamma _{{\rm{therm}}}^{{\bf{k'k}}}\left( n_{{\rm Pol}|{\bf k'}} + 1 \right){n_{{\rm Pol}|{\bf k}}} \right\} 
\end{multline}
\begin{multline} \label{NumberOfDarkmolecular vibrations}
\frac{d n_{\rm Vib_{D}}}{dt} 
=  
-  \gamma_{\rm Vib} \left( n_{\rm Vib_{D}} - n_{\rm Vib}^{\rm th}  \right) 
+ 
{ \gamma_{\rm Vib}^{\rm B-D} \over N_{\rm mol}}
\sum_{\bf k} 
\left(
\right.
\\
\left.
 n_{{\rm Vib}|{\bf k}_{\rm ex}-{\bf k}} - 
n_{\rm Vib_{D}} \right)
+ 
\sum_{\bf k}
 \frac{ G_{\bf k}}{N_{\rm mol}}  
\left[
n_{\rm Exc_{D}} \left(n_{{\rm Pol}|{\bf k}} + 
1 \right) 
+
\right.
\\
\left.
n_{\rm Vib_{D}} \left( n_{\rm Exc_{D}} - n_{{\rm Pol}|{\bf k}} \right) 
\right]
\end{multline}
\begin{multline} \label{NumberOfBrightmolecular vibrations}
\frac{d n_{{\rm Vib}|{\bf k}_{\rm ex}-{\bf k}}}{dt} 
=  
-  \gamma_{\rm Vib} \left( n_{{\rm Vib}|{\bf k}_{\rm ex}-{\bf k}} - n_{{\rm vib}}^{\rm th}  \right)
+
 \gamma_{\rm Vib}^{\rm B-D} 
\\
\left( n_{\rm Vib_{D}} - n_{{\rm Vib}|{\bf k}_{\rm ex}-{\bf k}} \right)
+ 
\frac{G_{\bf k}}{N_{\rm mol}} 
\left[
n_{\rm Exc|{\bf k}_{\rm ex}} \left(n_{{\rm Pol}|{\bf k}} + 1 \right) 
+
 \right.
\\
\left.
n_{{\rm Vib}|{\bf k_{\rm ex}}-{\bf k}} \left( n_{\rm Exc|{\bf k}_{\rm ex}} - n_{{\rm Pol}|{\bf k}} \right) 
\right]
\end{multline}
This includes polaritons $n_{{\rm Pol}|{\bf k}}$ and bright vibrational states $n_{{\rm Vib}|{\bf k}_{\rm Vib}}$, associated with momenta $\hbar\bf k$ and $\hbar\bf k_{\rm Vib}$ respectively, along with the occupation for the resonantly pumped bright exciton state $n_{{\rm Exc}|{\bf k}_{\rm ex}}$, all dark excitons $n_{\rm Exc_{D}}$, and vibrations $n_{\rm Vib_{D}}$. Hereinafter, we distinguish between the lower polariton states with relatively small in-plane momenta within $k_{B}T$ energy range around the ground state ${\hbar\bf k}={\bf 0}$ and polariton states with the large momentum ${\hbar\bf k}_{\rm ex}$, that we attribute to bright excitons. Indeed, lower polaritons at ${\hbar\mathbf{k}}_{\rm ex}$ have a dominant excitonic nature, unlike upper polaritons with the same momentum, which we disregard due to their large detuning from the resonant pumping condition $\hbar\omega_{{\rm Pump}}= \hbar\omega_{\rm Exc}$.
Here, $n_{{\rm Exc}|{\bf k}_{\rm ex}}$ refers to the number of dressed bright excitons pumped directly by the resonant excitation, $n_{\bf k}$ is the average number of lower polaritons in the state with $\hbar\bf k$ in-plane momentum, $n_{\rm Exc_{D}}$ represents the number of dressed dark excitons, $n_{\rm Vib}$ and $n_{\rm Vib_{D}}$ denote the number of bright and dark molecular vibrations, respectively. The term $n_{\rm Vib}^{\rm th}=(e^{\hbar\omega_{\rm Vib}/k_{B}T}-1)^{-1}$ is the vibrational population in the thermal equilibrium with an environment at temperature $T$. The parameters $\gamma_{\rm Exc}$, $\gamma_{{\rm Pol}|{\bf k}}$ and $\gamma_{\rm Vib}$ are the energy dissipation rates for excitons, lower polaritons with wavevector $\bf k$ and molecular vibrations respectively. Thermalization rates $\gamma_{{\rm{therm}}}^{\bf kk'}$ and $\gamma_{{\rm{therm}}}^{\bf k'k}$ represent downward and upward energy relaxation in the polariton subsystem, respectively, linked by the Kubo–Martin–Schwinger relation~\cite{kubo1957statistical, martin1959theory, breuer2002theory, kavokin2017microcavities}. Transition rates between bright and dark states $\gamma_{\rm Exc}^{\rm B-D}$ and $\gamma_{\rm Vib}^{\rm B-D}$ correspond to the relaxation of the bright excitonic and vibrational states to the manifold of localized dark states due to IVR mechanisms leading to the dephasing $\Gamma_{\rm Exc} = \gamma_{\rm Exc}^{\rm B-D}$ and $\Gamma_{\rm Vib} = \gamma_{\rm Vib}^{\rm B-D}$, respectively. The resonant pumping rate $\varkappa_{\rm Pump}$, being devided by total number of molecules $N_{\rm mol}$ has a clear physical meaning of the stationary excitation number per molecule.

\begin{figure}
\includegraphics[width=1\linewidth]{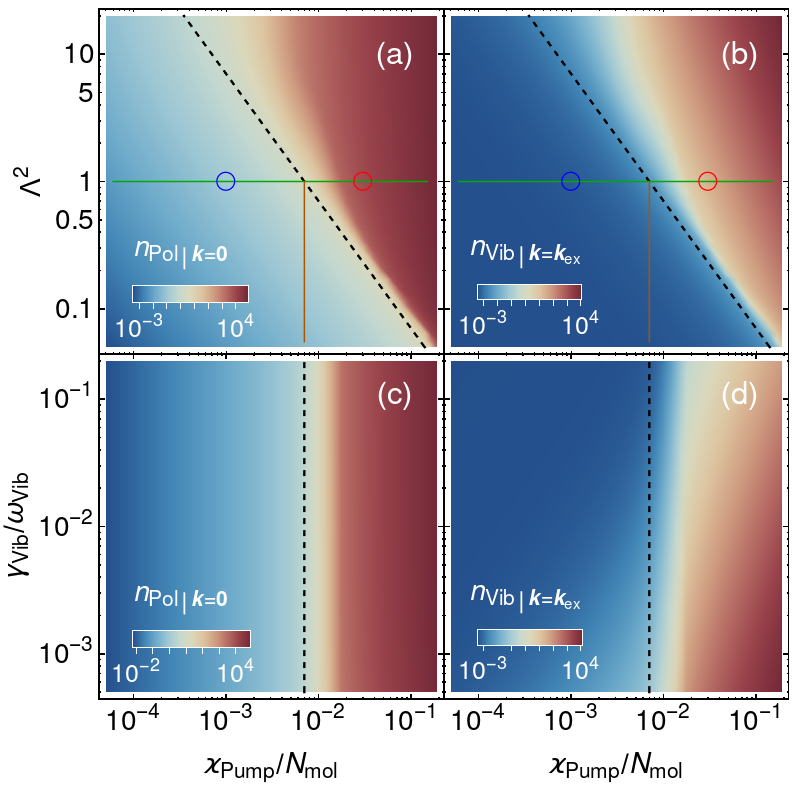}
\caption{Phase diagrams for polariton and vibrational states demonstrating the regime of coexisting polariton BEC and macroscopically occupied vibrational states. (a),(b) - Polariton and vibrational occupation numbers as the function of Huang--Rhys factor $\Lambda^{2}$ and the resonant pumping rate $\varkappa_{\rm Pump}$ normalized to the number of molecules $N_{\rm mol}$. Vibrational decoherence rate is assumed to be lifetime-limited $\Gamma_{\rm Vib} = \gamma_{\rm Vib}/2 = 5\times10^{-3} \omega_{\rm Vib}$. (c),(d) - Polariton and vibrational occupation numbers as the function of vibrational decoherence rate $\Gamma_{\rm Vib}$ and the resonant pumping rate $\varkappa_{\rm Pump}$. Here we fix Huang-Rhys factor $\Lambda^{2} = 1$. Black dashed line shows the analytical value of the threshold for polariton and vibrational macroscopically occupied states given by Eq.~(\ref{threshold}). The parameters here are $N_{\rm mol}=10^8$, $\Omega_R = 0.05$~$\rm eV$, $\gamma_{\rm Exc}=10^{-5}$~$\rm{eV}$, $\gamma_{\rm Exc}^{\rm B-D} = \Gamma_{\rm Exc} = 10^{-2}$~$\rm{eV}$,
$\gamma_{\rm Vib}^{\rm B-D} = \Gamma_{\rm Vib}$, 
$\gamma_{{\rm{therm}}}^{\bf kk'}= 10^{-5}$~$\rm eV$ for $|{\bf k}| < |{\bf k'}|$ and $T=290$~$\rm{K}$,
$\gamma_{{\rm Pol}|{\bf k}}\approx\gamma_{\rm {Cav}|{\bf{k}}}=2.5
\cdot10^{-3}$~$\rm{eV}$, $\omega_{{\rm Cav}|{\bf k}} = \omega_{{\rm Cav}|{\bf k}={\bf 0}} + \alpha_{\rm Cav} {\bf k}^2$ with $\omega_{{\rm Cav}|{\bf k}={\bf 0}} = 2.52~{\rm eV}$, $\alpha_{\rm Cav} = 2 \cdot 10^{-3}$~${\rm eV} \mu {\rm m^2}$, $\omega_{\rm Exc} = 2.72~{\rm eV}$, and $S = 500$~$\mu{\rm m^2}$. These parameters are consistent with the recent experiments~\cite{zasedatelev2019room, zasedatelev2021single}.
}
    \label{fig:Fig2}
\end{figure}

The constant $G_{\bf k}$ plays a central role in polariton condensation and generation of macroscopic vibrational states:
\begin{equation}\label{optomechanical_constant}
 G_{\bf k}
=
\frac{ \Lambda^2 \Omega_R^2 \Gamma_{\rm Exc} \cos^2\varphi_{\bf k}}{ (\omega_{\rm Exc}-\omega_{{\rm Pol}|{\bf k}}-\omega_{{\rm Vib}})^2 + (\Gamma_{\rm Exc}/2)^2 }
\end{equation}

\noindent where $\omega_{\rm Exc} = \omega_{\rm exc} - \Lambda^2 \omega_{\rm Vib}$ and $\Omega_R^2=\sum_{j=1}^{N_{\rm mol}}|\Omega_{j{\bf k}}|^2$ represents the total Rabi frequency. Serving as the effective polariton-vibration coupling constant, $G_{\bf k} $ characterizes the energy transfer rate
between exciton-polariton and vibrational states effectively describe by
the term 
$\sum_{{\bf k}'{\bf k}} \hbar g_{\bf k'k}  \hat a_{{\rm Pol}|{\bf k}}^\dag {\hat a_{{\rm Pol}|{\bf k}'}\left( {{{\hat b}_{{\rm Vib}|{\bf k}-{\bf k}'}} + b_{{{\rm Vib}|{\bf k}'-{\bf k}}}^\dag } \right)}$, where $g_{\bf k'k} = i(\Lambda \Omega_R/\sqrt{N_{\rm mol}})\sin(\varphi_{{\bf k}'}-\varphi_{{\bf k}})$~\cite{shishkov2024optomech}. 
Evidently, this coupling corresponds to a conventional 
multimode 
cavity optomechanical Hamiltonian~\cite{aspelmeyer2014cavity}, except that here, we deal with a hybrid light-matter state (exciton-polariton) rather than a bare cavity mode. 
The idea of using exciton-polariton states in cavity optomechanics have been originally developed for GaAs microcavities~\cite{kyriienko2014optomechanics, carlon2022enhanced, jusserand2015polariton, santos2023polaromechanics}. 
However, the low frequency of mechanical modes in these systems ($\sim$ GHz)~\cite{kuznetsov2023microcavity}, and the small exciton binding energy~\cite{deng2010exciton} make their operation in a coherent regime difficult to impossible under the high thermal population at room temperature. 
In our case, high-energy molecular vibrational modes take on the function of a mechanical oscillator. The optomechanical interaction steers population dynamics across the cavity and mechanical degrees of freedom depending on the optomechanical back-action rate $ \sim 
|g_{{\bf k}_{\rm ex}{\bf k}}|^{2} \times n_{{\rm Exc}|{\bf k}_{\rm ex}}
$~\cite{aspelmeyer2014cavity}. In our case, the blue-detuned external laser drive $\hbar\omega_{\rm Pump} = \hbar\omega_{{\rm Pol}|{\bf k} = {\bf 0}} + \hbar\omega_{\rm Vib}$ imposes a negative back action rate (anti-damping), resulting in strong vibrational amplification under certain conditions, as illustrated in Figure~\ref{fig:Fig1}.

Equations~(\ref{NumberOfBrightExcitons})--(\ref{NumberOfBrightmolecular vibrations}) offer a reduced microscopic model able to simulate dynamics of electronic and vibrational states of molecules with strong vibronic coupling in realistic cavities, given appropriate choice of experimental parameters. 
We numerically solve Eqs.~(\ref{NumberOfBrightExcitons})--(\ref{NumberOfBrightmolecular vibrations}) and find a steady-state population of polaritons at the ground state (${\bf k} = {\bf 0}$) as well as the population of the bright vibrational state mediating polariton condensation that fulfils phase matching condition $\hbar \bf k_{\rm Vib} = \hbar \bf k_{\rm Pump||} - \hbar \bf k_{\rm BEC} $ (see Ref.~\cite{shishkov2024optomech} for a discrete version of Eqs.~(\ref{NumberOfBrightExcitons})--(\ref{NumberOfBrightmolecular vibrations})). Figure~\ref{fig:Fig2} presents density plots of the polariton and vibrational occupation numbers as functions of the pumping rate $\varkappa_{\rm Pump} / N_{\rm mol}$, Huang-Rhys factor $\Lambda^{2}$, and vibrational decay rate $\gamma_{\rm Vib}$.
An analytic approximation to the steady-state solution reveals the threshold pumping rate as a function of polariton decay rate and optomechanical coupling (see SM~\cite{sm} Section II) 
\begin{equation} \label{threshold}
\varkappa_{\rm Pump}^{\rm thresh} 
= 
N_{\rm mol} 
\left.
{\gamma_{{\rm Pol}|{\bf k}} \over 
G_{{\bf k}}}
\right|_{{\bf k}={\bf 0}}. 
\end{equation}
Notably, the threshold value $\varkappa_{\rm Pump}^{\rm thresh} \propto \Lambda^{-2}$ does not depend on the vibrational decay rate  $\gamma_{\rm Vib}$, which makes our proposal robust against fast IVR mechanisms. 
Figures ~\ref{fig:Fig2}(a,c) show an average occupation number in the polariton subsystem as a function of Huang-Rhys factor and vibrational decay.

\begin{figure}
\includegraphics[width=1\linewidth]{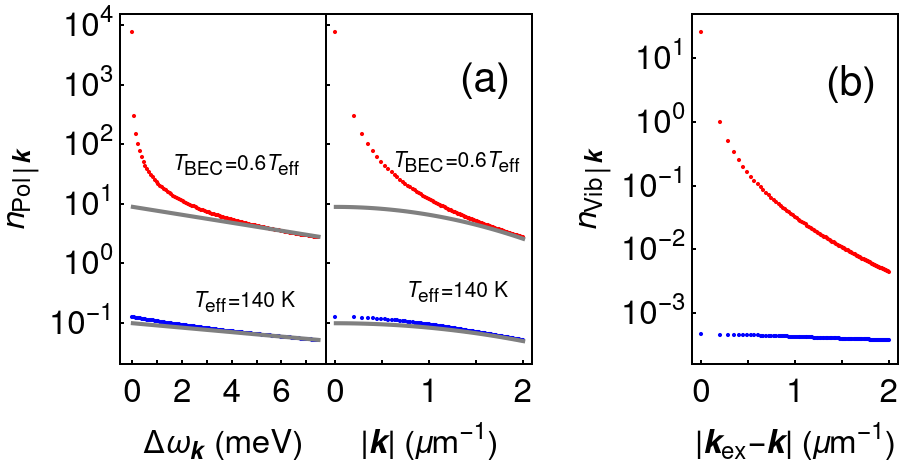}
\caption{
Energy (left) and momentum (right) distributions of polariton states below (blue) and above (red) condensation threshold - (a), where $\Delta\omega_{\bf k}=\omega_{{\rm Pol}|{\bf k}}-\omega_{{\rm Pol}|{\bf k}={\bf 0}}$. Momentum distribution of vibrational states below (blue) and above (red) the threshold - (b). The parameters are the same as used for Figure~\ref{fig:Fig2}a,b - indicated by the blue and red circles for below and above the threshold respectively.  Fitting curves in (a,b) represent the effective temperature determined by the Boltzmann distribution in energy and momentum spaces, respectively.
}
    \label{fig:Fig3}
\end{figure}


The formation of polariton BEC, in general, requires two conditions: 1 - the rate of polariton thermalization overcomes the energy dissipation ($\sum_{{\bf k}}\gamma_{\rm therm}^{{\bf k}{\bf 0}}n_{{\rm Pol}|{\bf k}} \gtrsim \gamma_{{\rm Pol}|{\bf k}}$), and 2 - the total number of lower polaritons surpasses the critical number~$\sum_{\bf k} n_{{\rm Pol}|{\bf k}}|_{\rm critical}\approx\nu k_B T$~\cite{shishkov2022exact,shishkov2022analytical}.
In our system, the first condition is met well below the threshold~(\ref{threshold}) while the second one is achieved at the threshold ${\sum_{\bf k} n_{{\rm Pol}|{\bf k}}|_{\rm thresh} \sim \sum_{\bf k} n_{{\rm Pol}|{\bf k}}|_{\rm critical} \sim 10^2-10^3}$ (see SM~\cite{sm} Section II). 
Followed by BEC formation of polaritons with in-plane wavevector $\bf k = \bf 0$, the system exhibits vibrational amplification towards macroscopically occupied vibrational state with ${\bf k}_{\rm Vib} = {\bf k}_{\rm ex}$ as shown Figure~\ref{fig:Fig2}b. Similar effect of phonon-amplification towards mechanical ``lasing'' enabled by strong polariton optomechanical interaction has been recently demonstrated in a GaAs microcavity cooled down to 5K~\cite{chafatinos2020cavity}. The matching threshold values for polariton and vibrational subsystems, unaffected by vibrational decay rate, highlights the fact that macroscopic occupation of the vibrational state is a result of polariton condensation.
This regime of coexisting macroscopically occupied exciton-polariton and vibrational states is a distinct feature of polariton systems with strong exciton-vibration interaction in the resonant blue-detuned optomechanical configuration: $\Delta_{\rm{OM}}=\omega_{\text{Pump}} - \omega_{\text{BEC}} = \omega_{\text{Vib}}$. Besides the optomechanical detuning it is worth mentioning the role of cavity detuning $\Delta_{\rm{Cav}}=\omega_{\rm{Cav}|\bf{k}=0}-\omega_{\rm{Exc}}$. As follows from Eq.8, red-detuned cavities $\Delta_{\rm{Cav}}<0$ are beneficial for the vibrational condensation, as they provide smaller polariton decay rate  defined by the cavity lifetime. Indeed, $\gamma_{{\rm Cav}}<\Gamma_{{\rm Exc}}$ in the most polariton systems. Obviously, the higher cavity quality factor ($Q$) the lower is the condensation threshold, see details in SM~\cite{sm} Section II.  
Although the decay of molecular vibrations does not explicitly influence the threshold, it does have an impact on the average vibration occupation number, as shown in Figure~\ref{fig:Fig2}(d) and is also evidenced by the relationship $n_{{\rm Vib}|{\bf k}_{\rm ex}} \propto \gamma_{\rm Vib}^{\rm -1}$.


The polariton optomechanical interaction in our system allows the state of the BEC to be mapped onto vibrational degrees of freedom. We consider two regimes: one below condensation threshold and the other above it, as indicated by the blue and red circles in Figures~\ref{fig:Fig2}a,b respectively. 
Figure~\ref{fig:Fig3}a illustrates energy and momentum distributions of polaritons below and above condensation threshold. Fast thermalization brings polaritons to dynamic equilibrium with an effective temperature $T_{\rm eff} \approx 140K$.
Above the threshold, the system undergoes a Bose--Einstein distribution with the thermalized tail at $T_{\rm BEC} = 0.6T_{\rm eff}$, a distinctive cooling effect that comes from the nonequilibrium nature of polariton BEC, recently demonstrated in Ref.~\cite{shishkov2022exact}. In momentum space, Figure~\ref{fig:Fig3}b, vibrational states lock to the polariton distribution due to resonant phase-matching condition. When the pumping rate exceeds the condensation threshold, it collapses to the bright state with a well-defined in-plane momentum $\hbar {\bf k}_{\rm ex}$. In Ref.\cite{shishkov2024optomech} we reveal that the optomechanical interaction with the polariton subsystem renders this state the lowest energy state among the modes in the momentum space, thereby constituting the ground state of the vibrational subsystem. Furthermore, the higher-lying states exhibit thermalized distribution resembling the distribution of the polariton BEC depicted in Figure~\ref{fig:Fig3}a. Essentially, the density-driven polariton condensation enable macroscopic occupation at the single vibrational state with the lowest energy and a distinct momentum followed by thermalized destribution for the states with other momenta. This process is reminiscent
to the sympathetic cooling via ultra-cold atoms~\cite{jockel2015sympathetic}, yet the vibrational condensation involves coupling mechanical degrees of freedom to the non-equilibrium exciton-polariton BEC. In addition to macroscopic occupation (i) and thermalization (ii), the vibrational state inherits coherence properties of the exciton-polariton condensate (iii). Evidence supporting buildup of the long-range order for molecular vibrations is presented in SM~\cite{sm} Section III.

\begin{figure}
\includegraphics[width=1\linewidth]{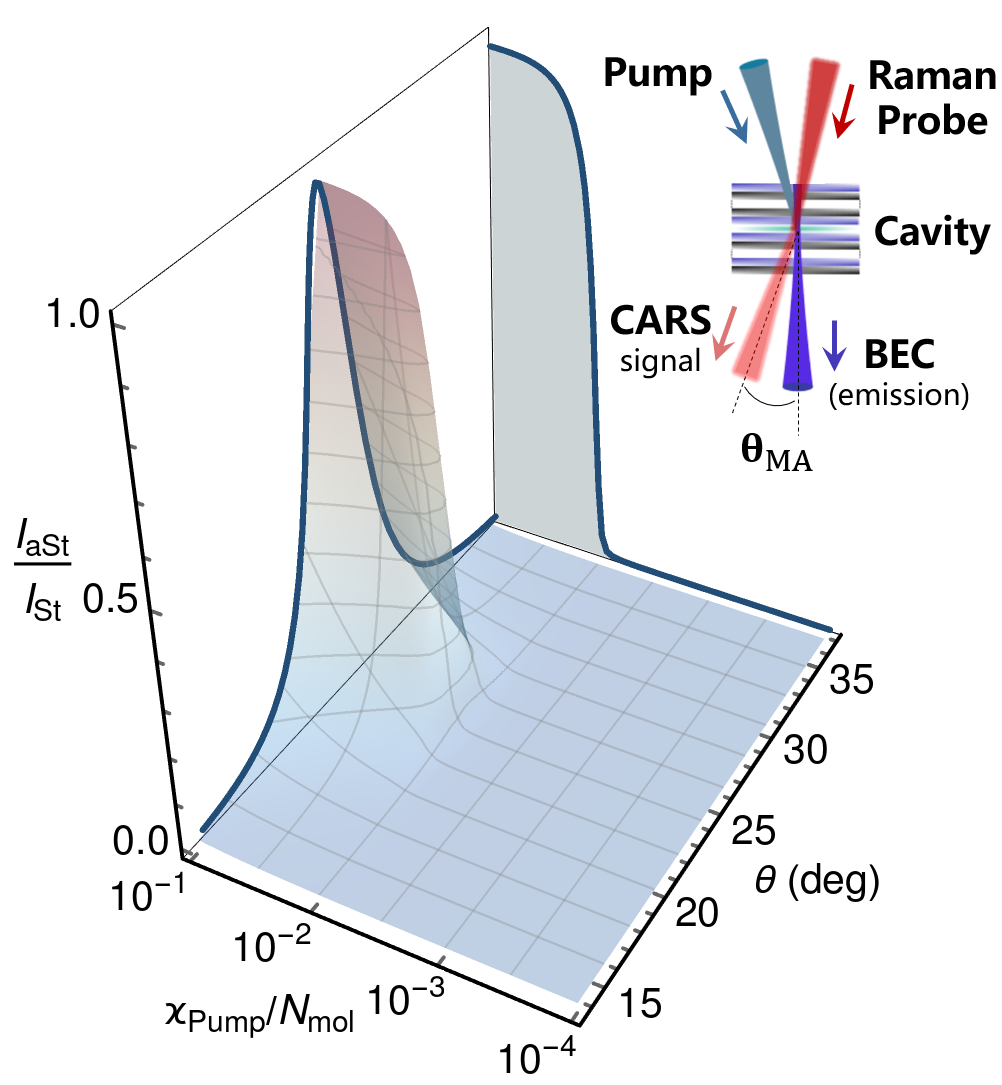}
\caption{Probing macroscopic vibrational states by non-resonant Raman spectroscopy of polariton BEC. The graph illustrates the angle-resolved and pump-dependent distribution of the intensity ratio between anti-Stokes and Stokes components of the Raman scattering, with parameters $\hbar \omega_{\rm Probe} = 2~{\rm eV}$, $\theta_{\rm Probe}=-30^{\rm o}$ and the rest parameters are the same as for Figure~\ref{fig:Fig2}. The inset shows proposed experimental arrangement.}
    \label{fig:Fig4}
\end{figure}

To validate the presence of macroscopically occupied vibrational states and its distribution, we propose angle-resolved non-resonant Raman spectroscopy of polariton BEC. The respective experimental arrangement is shown in Figure~\ref{fig:Fig4}. 
We calculate the intensity of the Stokes and anti-Stokes components of Raman scattering for a non-resonant probe beam with an energy $\hbar \omega_{\rm Probe}$ and in-plane momentum $\hbar \bf k_{\rm Probe}$. In the practical setups, the probe beam must be synchronized with the onset of polariton condensation. 
Figure~\ref{fig:Fig4} demonstrates the ratio between the anti-Stokes and Stokes components as a function of both pumping rate and detection angle. A coherent anti-Stokes Raman scattering (CARS) signal emerges at a specific ``magic angle'' (MA), denoted as $\theta_{\rm MA}$ (see Eq.~(\ref{magic angle})), due to resonant vibrational amplification.

\begin{equation} \label{magic angle}
\sin \theta_{\rm MA} 
=
{
\omega_{\rm Pump} \sin \theta_{\rm Pump} 
+ 
\omega_{\rm Probe} \sin \theta_{\rm Probe}
\over
\omega_{\rm Probe} + \omega_{\rm Vib}
}
\end{equation}

\noindent The magic angle provides a direct measurement of momentum for the vibrational mode $\hbar \bf k_{\rm ex}$ corresponding to the resonant condition (see SM~\cite{sm} Section IV for details). The formation of macroscopically occupied vibrational states is manifested by the CARS signal, which asymptotically approaches intensity of the Stokes component above condensation threshold, as depicted in Figure~\ref{fig:Fig4}.

In conclusion, we have introduced a new way of generating macroscopic vibrational states through exciton-polariton condensation in the strong vibronic regime. 
We have demonstrated that an effective strong optomechanical coupling between vibrational and polariton degrees of freedom can be achieved, characterized by interaction strength $G_{\bf k}$. This coupling results in parametric amplification of the resonant vibrational states above condensation threshold in close analogy to the cavity optomechanics in the blue-detuned configuration~\cite{aspelmeyer2014cavity}. 
The proposed configuration is consistent with practical high-$Q$ microcavities based on molecular systems having high-energy vibrational modes $\hbar\omega_{\rm Vib}\simeq200~{\rm meV}$ and featuring strong exciton-vibration interaction $\Lambda^{2}\sim1$, such as MeLPPP~\cite{plumhof2014room,zasedatelev2021single} or Anthracene~\cite{kena2008strong, kena2010room} polariton systems, for details see SM~\cite{sm} Section V.

We suggest a practical setup using a non-resonant Raman probe to identify macroscopic vibrational states coupled to the exciton-polariton BEC. 
Protected from natural decoherence by the cavity, coherence time in this regime is limited by nonlinear effects within the BEC~\cite{betzold2019coherence,yagafarov2020mechanisms} exceeding inherent vibronic decoherence~\cite{mukamel2000multidimensional,song2015separation} by at least three orders of magnitude at room temperature~\cite{betzold2019coherence,putintsev2020nano}. 
Through engineering a nonlinear potential landscape~\cite{askitopoulos2019giant,baryshev2022engineering}, it is possible to further extend coherence beyond nanoseconds, especially in advanced multi-component polariton systems~\cite{mcghee2023ultrafast,putintsev2023shaping}.
Therefore our study promise conceptual implications for understanding coherence in molecular systems and its compatibility with various chemical and physical processes~\cite{scholes2017using}. The massive occupation number of a single state at the BEC offers an elegant solution to the problem of large dark-state density, which often deters the efficiency of cQED effects in physical and chemical processes~\cite{vurgaftman2020negligible}. Bosonic stimulation at the BEC efficiently surpasses unwanted decay into manifold of the dark states, channeling the energy or enhancing reaction rates by a factor of $\sim (N+1)$ when the final/product state is at the BEC with occupation number $N$~\cite{phuc2022bose}.
Expanding upon the proposed scheme to include non-resonant vibrational control methods combined with recently developed methods of resonant polariton seeding~\cite{zasedatelev2019room,baranikov2020all} paves the way for manipulating light-matter states across a broad spectral range~\cite{shishkov2024optomech} for nonlinear and quantum optics with molecules at room temperature.

V.Yu.Sh.~thanks Foundation for the Advancement of Theoretical Physics and Mathematics "Basis" for financial support. A.V.Z. acknowledges support from the European Union's Horizon 2020 research and innovation programme under the Marie Skłodowska-Curie grant agreement No 101030987 (LOREN). This work was performed, in part, at the Center for Integrated Nanotechnologies, an Office of Science User Facility at Los Alamos National Laboratory operated for the U.S. Department of Energy (DOE) Office of Science.


\bibliography{main-vPRLwRefs}

\clearpage

\begin{widetext}

\section*{Supplementary Materials}

\section{Validity of the mean-field approximation}

The mean-field approach is valid both for the condensed part and for the non-condensed part of the system. We tested this by running simulations beyond the mean-field approach, using a recently developed quantum theory for the density matrix of an ideal Bose gas with pumping, dissipation, and fast thermalization~\cite{shishkov2022exact}. The mean-field theory applied to this case of a nonequilibrium ideal Bose gas returns the same result~\cite{banyai2002real} for the population of non-condensed polaritons above the condensation threshold as the fully quantum theory predicts~\cite{shishkov2022exact}. Moreover, the theory agrees well with experimental data, reproducing the Bose-Einstein distribution for the population of higher-energy, above-condensate polariton states~\cite{zasedatelev2021single}, which further highlights the mean-field approach as the relevant approximation.

Nevertheless, we believe that future developments beyond mean-field theories are important for understanding quantum correlations, especially within the build-up phase of polariton BEC. In this regard, the approach based on the truncated Wigner approximation~\cite{phuc2024semiclassical} stands as a promising theory to gain insights to the quantum statistics of the polariton and vibrational states, as well as their mutual correlations and possible entanglement. These aspects are beyond the scope of our work, but are of great interest for future investigation.

\section{\label{sec:Threshold} Analytic approximation to the condensation threshold}
To facilitate further derivation of the analytic approximation we introduce the following notations
\begin{equation}
I = n_{{\rm Exc}|{\bf k}_{\rm ex}} + N_{\rm mol} n_{\rm Exc_D},
\end{equation}
\begin{equation}
P = \sum_{\bf k} n_{{\rm Pol}|{\bf k}}.
\end{equation}
\begin{equation}
V = N_{\rm mol} n_{\rm Vib_D} + \sum_{\bf k} n_{{\rm Vib}|{\bf k}},
\end{equation}

From the Eqs.~(2)--(6) in the main text we obtain
\begin{equation} \label{eq I}
{d I \over dt} 
= 
- \gamma_{\rm Exc} (I - \varkappa_{\rm Pump}) 
- {1 \over N_{\rm mol}} I (P G_{{\bf k}={\bf 0}} + G_S)
- {1 \over N_{\rm mol}} J
\end{equation}
\begin{equation} \label{eq P}
{d P \over dt}
=
- \gamma_{{\rm Pol}|{\bf k}={\bf 0}} P
+ {1 \over N_{\rm mol}} I (P G_{{\bf k}={\bf 0}} + G_S)
+ {1 \over N_{\rm mol}} J
\end{equation}
\begin{equation} \label{eq V}
{d V \over dt}
=
- \gamma_{\rm Vib} (V - N_{\rm mol} n_{\rm Vib}^{\rm th})
+ {1 \over N_{\rm mol}} I (P G_{{\bf k}={\bf 0}} + G_S)
+ {1 \over N_{\rm mol}} J
\end{equation}
were we assume $\tilde G_{\bf k} \approx G_{\bf k}$, $\gamma_{{\rm Pol}|{\bf k}_{\rm ex}} \approx \gamma_{\rm Exc}$, $\sum_{\bf k} G_{\bf k} n_{{\rm Pol}|{\bf k}} \approx G_{{\bf k}={\bf 0}} P$, $\sum_{\bf k} \gamma_{{\rm Pol}|{\bf k}} n_{{\rm Pol}|{\bf k}} \approx \gamma_{{\rm Pol}|{\bf k}={\bf 0}} P$, $\sum_{\bf k} G_{\bf k} n_{{\rm Vib}|{\bf k}_{\rm ex}-{\bf k}} \approx G_{{\bf k}={\bf 0}} V$, $N_{\rm mol} + \sum_{\bf k} 1 = N_{\rm mol} + N_{\rm states} \approx N_{\rm mol}$
and denote
\begin{equation}
J = \sum_{\bf k} G_{\bf k} 
\left[
n_{{\rm Vib}|{\bf k}_{\rm ex}-{\bf k}}( n_{{\rm Exc}|{\bf k}_{\rm ex}} - n_{{\rm Pol}|{\bf k}} )
+ N_{\rm mol} n_{\rm Vib_D} (n_{\rm Exc_D}-n_{{\rm Pol}|{\bf k}})
\right],
\end{equation}
\begin{equation}
G_S=\sum_{\bf k} G_{\bf k}.
\end{equation}

Here we consider a situation when the number of vibrations per molecule is much less than one.
In this case, the term $J/N_{\rm mol}$ is much smaller than the term $I (P G_{{\bf k}={\bf 0}} + G_S)/N_{\rm mol}$ in the Eqs.~(\ref{eq I})--(\ref{eq V}).
Therefore we can neglect this term and obtain the stationary solution for the total amount of polaritons
\begin{equation}
\sum_{\bf k} n_{{\rm Pol}|{\bf k}}
=
{\gamma_{\rm Exc} \over 2\gamma_{{\rm Pol}|{\bf k}={\bf 0}}} (\varkappa_{\rm Pump} - \varkappa_{\rm Pump}^{\rm thresh})
+
\sqrt{
\left[
{\gamma_{\rm Exc} \over 2\gamma_{{\rm Pol}|{\bf k}={\bf 0}}} (\varkappa_{\rm Pump} - \varkappa_{\rm Pump}^{\rm thresh})
\right]^2
+
\varkappa_{\rm Pump} {\gamma_{\rm Exc} \over \gamma_{{\rm Pol}|{\bf k}={\bf 0}}} {G_S \over G_{{\bf k}={\bf 0}}}
},
\end{equation}
where the threshold pumping rate is defined as
\begin{equation} \label{threshold}
\varkappa_{\rm Pump}^{\rm thresh} 
= 
N_{\rm mol} {\gamma_{{\rm Pol}|{\bf k}={\bf 0}} \over G_{{\bf k}={\bf 0}}} 
+
{G_S \over G_{{\bf k}={\bf 0}}} {\gamma_{{\rm Pol}|{\bf k}={\bf 0}} \over \gamma_{\rm Exc}}
\end{equation}
For the system considered in the main text, the second term of the right-hand side of the Eq.~(\ref{threshold}) is much smaller than the first one.
Therefore, in the main text, we preserve only the first term.

We present the analysis of the Eq.~(\ref{threshold}) in the main text.

\begin{figure*}
\includegraphics[width=0.9\linewidth]{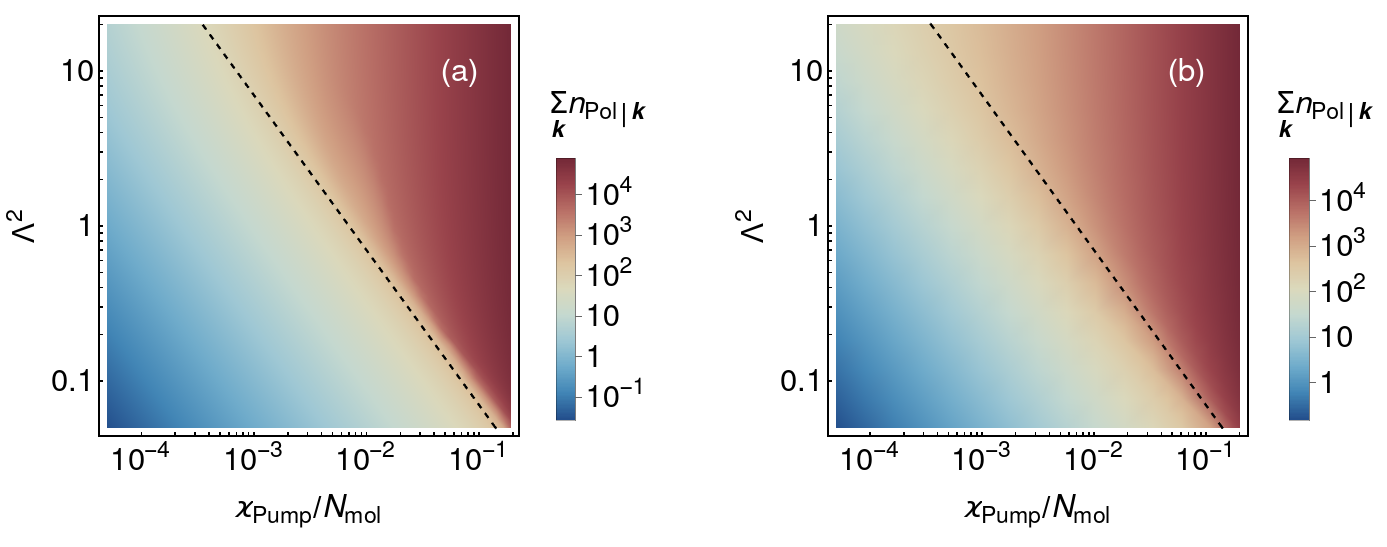}
\caption{Numerical simulations (a) and analytic results (b) obtained for the total number of lower polaritons.
The dashed black line shows the analytic approximation to the threshold values given by Eq.~(\ref{threshold}).
}
    \label{fig:analytics}
\end{figure*}

\begin{figure*}
\includegraphics[width=1\linewidth]{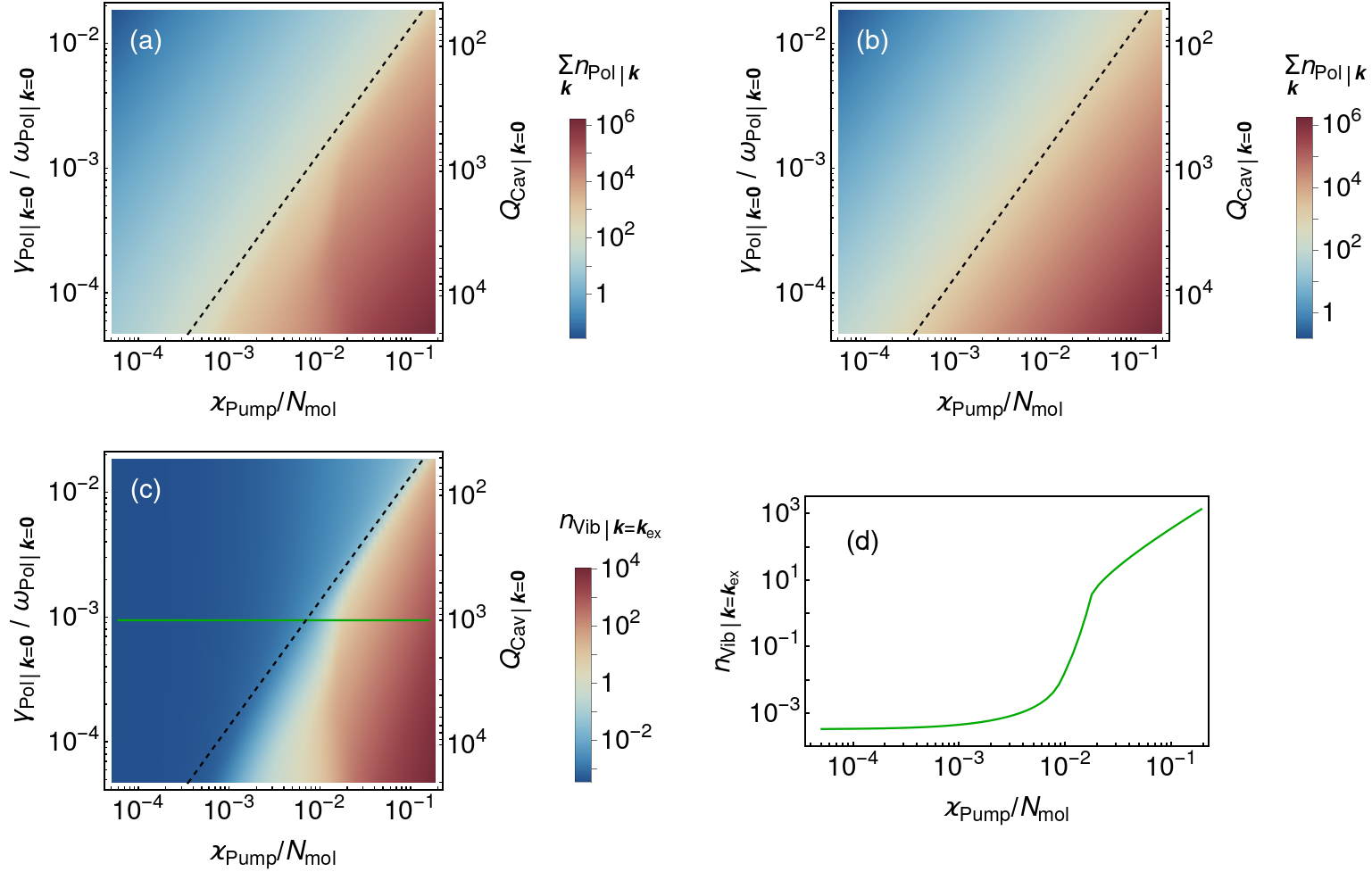}
\caption{Numerical simulations (a) and analytic results (b) obtained for the total number of lower polaritons, (c) numerical results for bright molecular vibrations with wave vector ${\bf k}_{\rm ex}$, (d) numerical results for bright molecular vibrations with wave vector ${\bf k}_{\rm ex}$ for $Q_{\rm cav}=10^{3}$.
The dashed black line shows the analytic approximation to the threshold values given by Eq.~(\ref{threshold}).
}
    \label{fig:analytics2}
\end{figure*}

On Figure~\ref{fig:analytics} and~\ref{fig:analytics2}  we compare the numerical simulations for $\sum_{\bf k} n_{{\rm Pol}|{\bf k}}$ and the analytic result~Eq.~(\ref{analytics}).
One can see, that analytic approximation is in good agreement with the numerical simulations, showing that higher Huang-Rhys factor and higher quality factor of the cavity lower the threshold.
The values $\sum_{\bf k} n_{{\rm Pol}|{\bf k}}$ at the threshold exhibit an abrupt change, a feature that is more pronounced when comparing numerical simulations to analytics.
We believe this discrepancy comes from our analytic approach where we neglect the term $J/N_{\rm mol}$ in the Eqs.~(\ref{eq I})--(\ref{eq V}).
This term, in particular, describes the influence of the microscopic occupation of the bright vibrational states on the BEC dynamics.


At the threshold~(\ref{threshold}) the total number of the lower polaritons is given by
\begin{equation} \label{analytics}
\sum_{\bf k} n_{{\rm Pol}|{\bf k}}|_{\rm thresh}
=
\sqrt{
\varkappa_{\rm Pump}^{\rm thresh} {\gamma_{\rm Exc} \over \gamma_{{\rm Pol}|{\bf k}={\bf 0}}} {G_S \over G_{{\bf k}={\bf 0}}}
}
.
\end{equation}

The formation of a polariton BEC generally requires that the total number of lower polaritons surpasses the critical number below
\begin{equation}
\sum_{\bf k} n_{{\rm Pol}|{\bf k}}|_{\rm critical}\approx\nu k_B T
\end{equation}
where $\nu$ is the density of states for polaritons with the dispersion $\omega_{{\rm Pol}|{\bf k}} = \omega_{{\rm Pol}|{\bf k}={\bf k}} + \alpha_{\rm Pol} {\bf k}^2$
\begin{equation}
\nu \approx {S \over 4 \pi \alpha_{\rm Pol}}.
\end{equation}

For the realistic parameters of the system, considered in the main text we get the following total polariton number 
\begin{equation}
\sum_{\bf k} n_{{\rm Pol}|{\bf k}}|_{\rm thresh} 
\sim 
\sum_{\bf k} n_{{\rm Pol}|{\bf k}}|_{\rm critical} 
\sim 
10^2-10^3.
\end{equation}
Thus, crossing of the threshold~(\ref{threshold}) is immediately followed by the BEC formation.

Interestingly, $\sum_{\bf k} n_{{\rm Pol}|{\bf k}}|_{\rm thresh} \propto \Lambda^{-1}$ and $\sum_{\bf k} n_{{\rm Pol}|{\bf k}}|_{\rm critical}$ does not depend on $\Lambda$, whereas $\sum_{\bf k} n_{{\rm Pol}|{\bf k}}|_{\rm critical} \propto T$ and $\sum_{\bf k} n_{{\rm Pol}|{\bf k}}|_{\rm thresh}$ does no depend on $T$, given that $\Gamma_{\rm Exc}$ remains constant.
Therefore, by adjusting the Huang-Rhys factor and the temperature of the sample, one can distinguish between two regimes: the polariton laser and the BEC.

\section{Long-range order above the condensation threshold}
Off-diagonal long-range order above the condensation threshold is a distinctive property of the BEC~\cite{kasprzak2006bose, deng2010exciton}.
Although the coherence of electronic states in such nonequilibrium systems has been both theoretically~\cite{shishkov2022exact} and experimentally demonstrated~\cite{plumhof2014room, zasedatelev2021single}, the coherence of vibrational states has not yet been explored. 

We assess the long-range order using the first-order spatial coherence function $g^{(1)}_{\rm Pol}({\bf r})$ calculated according to~\cite{sarchi2007long, doan2008coherence}. This approach is widely acknowledged in the community and is considered a benchmark signature in the analysis of light-matter condensation.

$$
g^{(1)}_{\rm Pol}({\bf r})
=
\frac{\langle \hat\psi_{\rm Pol}^\dag({\bf 0})  \hat\psi_{\rm Pol}({\bf r}) \rangle}{\sqrt{\langle \hat\psi_{\rm Pol}^\dag({\bf 0})  \hat\psi_{\rm Pol}({\bf 0}) \rangle} \sqrt{\langle \hat\psi_{\rm Pol}^\dag({\bf r})  \hat\psi_{\rm Pol}({\bf r}) \rangle}}
$$
where $\hat \psi_{\rm Pol}({\bf r}) $ is an annihilation operator for polaritons at the position $\bf r$ with in-plane expansion
$$
\hat \psi_{\rm Pol}({\bf r}) = \sum_{\bf k} \hat s_{{\rm Pol}|{\bf k}} e^{i {\bf k} {\bf r}},
$$
and $\hat s_{{\rm Pol}|{\bf k}}$ -- is the annihilation operator for polaritons with the in-plane momenta $\hbar\bf k$.
Upon substitution of the operator $\hat \psi_{\rm Pol}({\bf r}) $ into the polariton first-order coherence function $g^{(1)}_{\rm Pol}({\bf r})$ above, we arrive at the following expression:
$$
g^{(1)}_{\rm Pol}({\bf r})
=
\frac
{\sum_{\bf k}n_{{\rm Pol}|{\bf k}}e^{i{\bf k}{\bf r}}}
{\sum_{\bf k}n_{{\rm Pol}|{\bf k}}},
$$
Extending this approach for the bright vibrational states one get analogous expression for the vibrational coherence:
$$
g^{(1)}_{\rm Br.Vib}({\bf r})
=
\frac{\langle \hat\psi_{\rm Br.Vib}^\dag({\bf 0})  \hat\psi_{\rm Br.Vib}({\bf r}) \rangle}{\sqrt{\langle \hat\psi_{\rm Br.Vib}^\dag({\bf 0})  \hat\psi_{\rm Br.Vib}({\bf 0}) \rangle} \sqrt{\langle \hat\psi_{\rm Br.Vib}^\dag({\bf r})  \hat\psi_{\rm Br.Vib}({\bf r}) \rangle}}
$$
where $\hat \psi_{\rm Br.Vib}({\bf r}) $ is the annihilation operator for bright vibrational states at the position  $\bf r$
$$
\hat \psi_{\rm Br.Vib}({\bf r}) = \sum_{\bf k} \hat c_{{\rm Vib}|{\bf k}} e^{i {\bf k} {\bf r}},
$$
and $\hat c_{{\rm Vib}|{\bf k}} $ is the annihilation operator for bright vibrational states with wave vector $\bf k$.
Therefore, the first-order spatial coherence function for vibrational states is defined as follows 
$$
g^{(1)}_{\rm Br.Vib}({\bf r}) 
=
\frac
{\sum_{\bf k}n_{{\rm Vib}|{\bf k}}e^{i{\bf k}{\bf r}}}
{\sum_{\bf k}n_{{\rm Vib}|{\bf k}}}.
$$

Figure~\ref{fig:g1} explicitly demonstrates the long-range order as $g^{(1)}({\bf r})$ for both exciton-polariton and vibrational states. It is evident that the vibrational subsystem inherits coherence properties of exciton-polariton states. This result is not very surprising though, considering that that the optomechanical amplification is a parametric process that preserves phase coherence between the involved excitonic and vibrational states. The schematic picture in Figure 1 of the main text shows the two coherent drives acting on the vibrational mode above condensation threshold: the laser pump and polariton BEC. Similar to coherent Raman scattering~\cite{song2015separation}, both waves map out their coherence onto the vibrational state. Consequently, vibrational states inherit the coherence properties of the nonequilibrium exciton-polariton BEC, subject to the nonlinear effects at the condensate~\cite{betzold2019coherence, yagafarov2020mechanisms}. Unlike conventional Raman-based methods, this enables concurrent electronic and vibrational states with significantly longer coherence at room temperature under narrow-band quasi-steady-state pumping conditions~\cite{putintsev2020nano, betzold2019coherence}.

\begin{figure}
\includegraphics[width=0.9\linewidth]{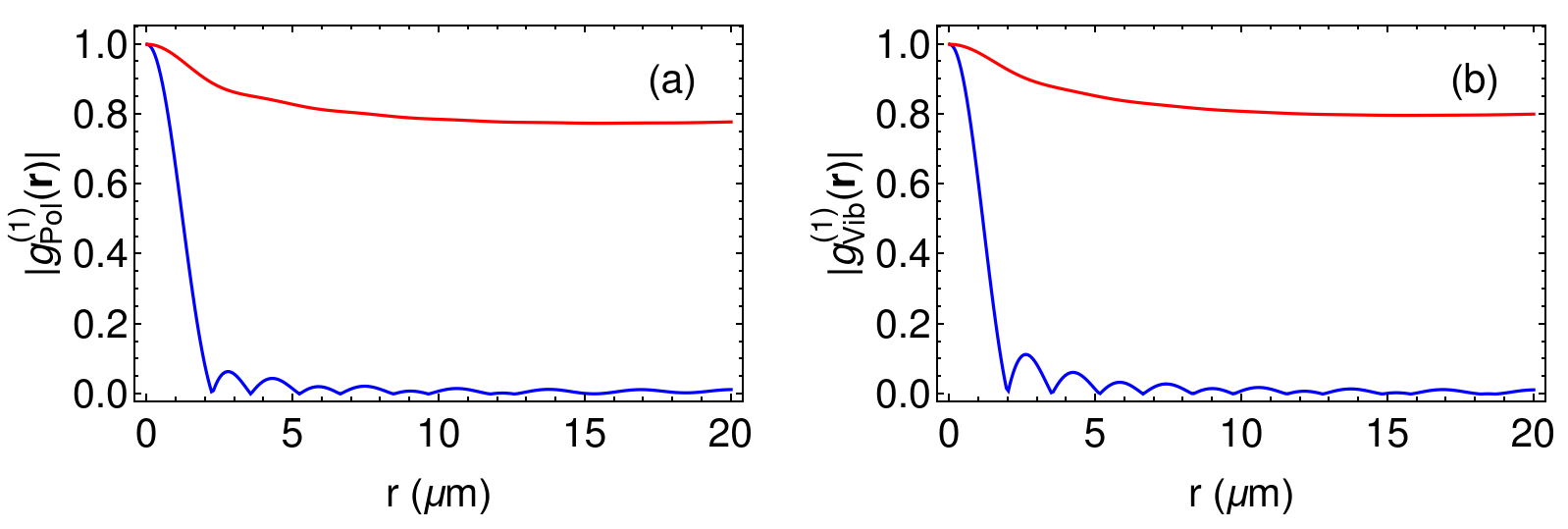}
\caption{The first-order spacial coherence function $g^{(1)}({\bf r})$ for polaritons - (a), and vibrations - (b). The parameters are the same as used for Figure 2 in the main text. The blue and red plots correspond the blue and red circles on Figure 2 in the main text for below and above the threshold respectively.
} \label{fig:g1}
\end{figure}

\section{\label{sec:CARS} Coherent anti-Stokes Raman Scattering on polariton BEC: magic angle conditions}

\begin{figure*}
\includegraphics[width=0.7\linewidth]{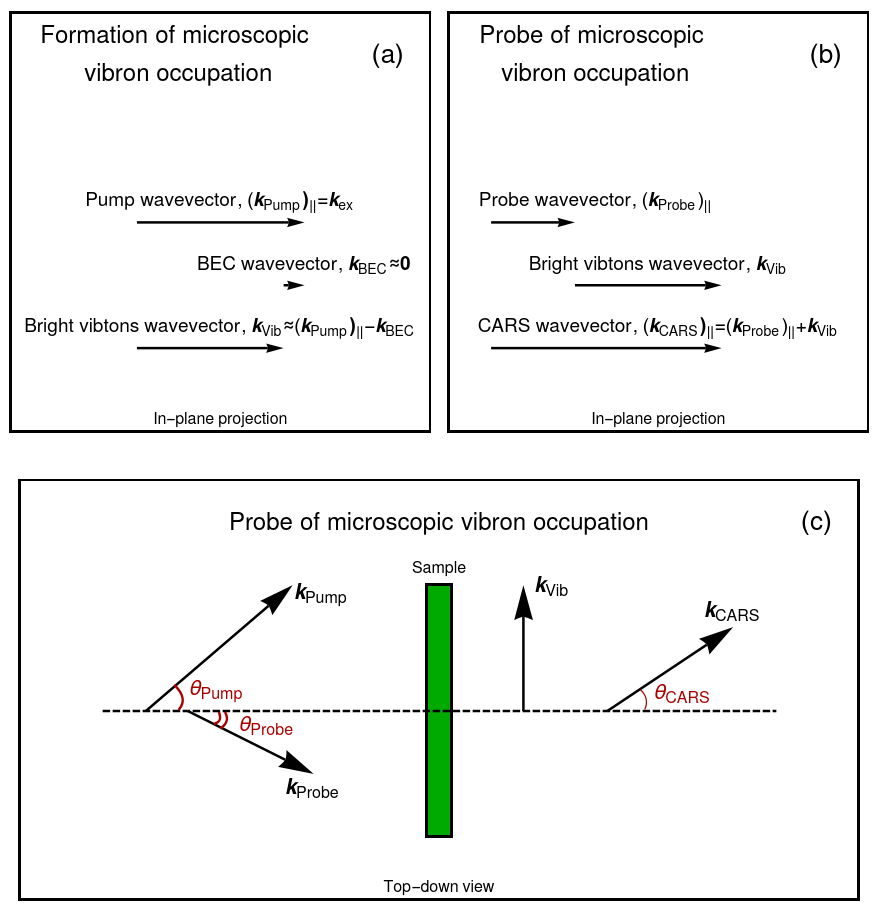}
\caption{Phase matching condition for the formation of polariton BEC and macroscopically occupied vibrational states (a), phase matching condition for the Raman probe (b) and schematic layout of the CARS measurement (c).
}
    \label{fig:CARS}
\end{figure*}

Here we calculate the magic angle of the CARS signal induced by Raman scattering from polariton BEC.
We consider the probe beam with the frequency $\omega_{\rm Probe}$ and the incident angle $\theta_{\rm Probe}$.
We denote the frequency of the pumping beam as $\omega_{\rm Pump}$ and its incident angle as $\theta_{\rm Pump}$.
We use the following assumption $\omega_{\rm Pump} = \omega_{\rm Exc}$.
Given the in-plane wavevector of the bright vibrations ${\bf k}_{\rm Vib}$ and in-plane wavevector of BEC ${\bf k}_{\rm BEC}$ we can express the energy conservation law in the following form
\begin{equation} \label{pump_w}
 \omega_{\rm Pump} = \omega_{\rm BEC} + \omega_{\rm Vib}
\end{equation}
\begin{equation} \label{probe_w}
 \omega_{\rm CARS} = \omega_{\rm Probe} + \omega_{\rm Vib}
\end{equation}
as well as in-plane phase matching condition
\begin{equation} \label{pump_k}
 ({\bf k}_{\rm Pump})_{||} = ({\bf k}_{\rm BEC})_{||} + {\bf k}_{\rm Vib}
\end{equation}
\begin{equation} \label{probe_k}
 ({\bf k}_{\rm CARS})_{||} = ({\bf k}_{\rm Probe})_{||} + {\bf k}_{\rm Vib}
\end{equation}
where $||$ denotes an in-plane component of the corresponding wavevector and $({\bf k}_{\rm Pump})_{||} = {\bf k}_{\rm ex}$.
Under the assumption that all vectors $({\bf k}_{\rm Probe})_{||}$, ${\bf k}_{\rm Vib}$ and $({\bf k}_{\rm Pump})_{||}$ lie in the same plane, one can treat them as c-numbers. Consequently, the following angles $\theta_{\rm Probe}$, $\theta_{\rm Pump}$ and $\theta_{\rm CARS}$ can be introduced (see Figure~\ref{fig:CARS})
\begin{equation} \label{projection_pump}
({\bf k}_{\rm Pump})_{||} = { \omega_{\rm Pump} \over c } \sin\theta_{\rm Pump}
\end{equation}
\begin{equation} \label{projection_probe}
({\bf k}_{\rm Probe})_{||} = { \omega_{\rm Probe} \over c } \sin\theta_{\rm Probe}
\end{equation}
\begin{equation} \label{projection_CARS}
({\bf k}_{\rm CARS})_{||} = { \omega_{\rm CARS} \over c } \sin\theta_{\rm CARS}
\end{equation}

The bright vibrational state with maximal occupation corresponds to the wavevector ${\bf k}_{\rm Vib} = {\bf k}_{\rm ex} = ({\bf k}_{\rm Pump})_{||}$, in agreement with Eq.~(\ref{pump_k}) where ${\bf k}_{\rm BEC} \approx {\bf 0}$. In the specific configuration of the polariton BEC we denote the corresponding angle $\theta_{\rm CARS}$ as $\theta_{\rm MA}$ and refer to it as the ``magic angle''.
Based on Eqs.~(\ref{pump_w})--(\ref{projection_CARS}) and condition ${\bf k}_{\rm BEC} = {\bf 0}$ we derive the expression for $\theta_{\rm MA}$ as presented in Eq.~(9) of the main text.

\section{Relevance to practical organic microcavities}
Without naming a particular chromophore explicitly we undertook calculation with the parameters relevant to the recent experiments demonstrating polariton condensation via the vibrational mechanism in MeLPPP-based microcavities~\cite{zasedatelev2019room,zasedatelev2021single}.

Our analysis includes strong exciton-vibration coupling, characterized by a Huang-Rhys factor of $\Lambda^{2}=1$, to the vibrational mode $\hbar\omega_{\rm Vib}=200~{\rm meV}~$\cite{guha2003temperature}. We employed realistic cavity parameters: a quality factor ($Q$) of 1000, Rabi splitting of $2\hbar\Omega_{R}=100~{\rm meV}$, and a negative cavity detuning ($\hbar\Delta_{\rm{Cav}}=\hbar\omega_{\rm{Cav}}-\hbar\omega_{\rm{Exc}}\simeq-200~{\rm meV}$)~\cite{plumhof2014room,zasedatelev2019room}, to achieve the necessary blue-detuned configuration $\Delta_{\rm{OM}}=\omega_{\text{Pump}} - \omega_{\text{BEC}} = \omega_{\text{Vib}} $ for the resonant excitation $\hbar\omega_{\text{Pump}}=\hbar\omega_{\text{Exc}}$, as shown in Figure 1 main text. 

These parameters closely align with those of Anthracene-based microcavities, which the referee mentioned. Such microcavities are known for their $Q\simeq800$, $2\hbar\Omega_{R}\simeq100~{\rm meV}$\cite{kena2008strong, kena2010room}, and demonstrate strong exciton-vibration coupling with a Huang-Rhys factor of $\Lambda^{2}=0.4$ to the vibrational mode of $\hbar\omega_{\rm Vib}=170~{\rm meV}$~\cite{coropceanu2002hole}, showing good quantitative agreement with our settings.

\end{widetext}


\end{document}